\documentclass[%
reprint,%
 amssymb, amsmath,%
 aip,jcp,%
groupedaddress,%
]{revtex4-1}

\usepackage[colorlinks=true,linkcolor=blue]{hyperref}%
\usepackage{graphicx}% Include figure files
\usepackage{dcolumn}% Align table columns on decimal point
\usepackage{bm}% bold math

\expandafter\ifx\csname package@font\endcsname\relax\else
 \expandafter\expandafter
 \expandafter\usepackage
 \expandafter\expandafter
 \expandafter{\csname package@font\endcsname}%
\fi
\newcommand{\lla}{\left\langle}
\newcommand{\rra}{\right\rangle}

\graphicspath{{./figures/}}

\begin{document}

%\preprint{AIP/123-QED}

\title{Active Polar Ring Polymer in Shear Flow -- An Analytical Study}

\author{Roland G. Winkler}
\email{rg\_winkler@gmx.de}
\affiliation{Theoretical Soft Matter and Biophysics, Institute for Advanced Simulation, Forschungszentrum J\"ulich, D-52425 J\"ulich, Germany}

\author{Sunil P. Singh}
\email{spsingh@iiserb.ac.in}
\affiliation{Department of Physics, Indian Institute of Science Education and Research,  Bhopal, Bhopal 462 066, Madhya Pradesh, India} 

\date{\today}% It is always \today, today,
             %  but any date may be explicitly specified

\begin{abstract}
We theoretically study the conformational and dynamical properties of semiflexible active polar ring polymers under linear shear flow. A ring is described as a continuous Gaussian polymer with a tangential active force of a constant density along its contour. The linear but non-Hermitian equation of motion is solved using an eigenfunction expansion, which yields activity-independent, but shear-rate-dependent, relaxation times and activity-dependent frequencies. As a consequence, the ring's stationary-state properties are independent of activity, and its conformations as well as rheological properties are equal to those of a passive ring under shear. The presence of characteristic time scales by the relaxation and the frequency gives rise to a particular dynamical behavior. A tank-treading-like motion emerges for large relaxation times and high frequencies, specifically for stiffer rings, governed by the activity-dependent frequencies. In the case of very flexible polymers, the relaxation behavior dominates over tank-treading. Shear strongly affects the crossover from a tank-treading to a relaxation-time dominated dynamics and suppresses tank-treading. This is reflected in the tumbling frequency, which exhibits two shear-rate dependent regimes, with an activity-dependent plateau at low shear rates followed by a power-law regime with increasing tumbling frequency for large shear rates. 
%This reflects the presence of two types of motion in shear flow for active ring polymers, dominance of tank-treading at small shear rates and tumbling at large ones. 

\end{abstract}

%\pacs{Valid PACS appear here}% PACS, the Physics and Astronomy
                             % Classification Scheme.
%\keywords{Suggested keywords}%Use showkeys class option if keyword
                              %display desired
\maketitle

\section{Introduction} 

Active soft matter consists of autonomously moving entities facilitated by their consumption of internal energy or energy from the environment to maintain an out-of-equilibrium state.\cite{cate:11,eise:16,doos:18,das:20,qi:22,joan:23} Particular examples are filaments and polymers of biological cells, with their out-of-equilibrium conformational and dynamical properties that determine the cells biological functions. \cite{nedl:97,das:20,wink:20}  Molecular motors walking along actin and microtubule filaments cause filament motion, which is fundamental for various cellular processes, including cell crawling, cell division, and chromosome segregation.\cite{nedl:97,milo:15,gudi:21} 

In vitro experiments on actin and microtubule filaments driven by molecular motors reveal bundle formation, a complex dynamics, and the emergence of active turbulence,\cite{nedl:97,howa:01,juel:07,hara:87,scha:10,sanc:12,sumi:12,pros:15,doos:18} emphasising the rich features of active polymeric systems. In particular, ring-like structures are formed,\cite{kawa:08,liu:11,keya:20,kawa:23} which exhibit a propulsion-driven rotational motion.\cite{kawa:08,keya:20,phil:22} 

In addition to propulsion, polymers and filaments in solution are exposed to nonequilibrium forces by flow, specifically in dilute solutions. Common to flow environments are gradients in fluid velocity such as, e.g., in linear shear flow. The appearing torques on the polymers/filaments, when coupled with motility, severely affect the polymer/filament conformations and dynamics, with fundamental consequences on their transport properties.

To shed light onto the emergent features of such out-of-equilibrium polymers, we analytically study the characteristics of tangentially driven active polar ring polymers (APRPs) under shear flow. Previously, the non-equilibrium properties of passive ring polymers under shear flow, particularly their conformations, have been analyzed in various simulations.\cite{mesl:09,chen:13.2,lang:14,lieb:18.1,fari:24} These studies provide insight into the particularities of ring geometry compared to linear polymers, specifically the emergence of tank-treading motion\cite{kell:82,lieb:18.1} and its dependence on more complex ring geometries.\cite{lieb:18.1,fari:24} In addition, the conformational and dynamical properties of self-propelled ring polymers comprised of active Brownian monomers (ABRPs) have been studied by simulations\cite{thee:24} and analytical theory,\cite{mous:21} revealing a strong influence of activity on the ring conformations -- swelling of flexible rings, and an initial shrinkage and latter swelling of semiflexible rings with increasing activity -- as well as an activity-enhanced dynamics. In an alternative approach, activity is modeled by a local  thermostat acting on monomers.\cite{smre:20} Here, attention has been paid to the influence of active on glass formation. Moreover, polar ring polymers combined with radially long-range isotropic forces along the polymer backbone have been studied by computer simulations and the combined effect of the ring conformational properties have been analyzed.\cite{kuma:23.1} The collective structural and dynamical properties of APRPs under confinement have also been addressed.\cite{mira:24} Recently, the combined influence of polar activity and shear flow on conformational and dynamical characteristics of ring polymers has been investigated, in particular, different bond potentials have been considered and differences revealed.\cite{kuma:24} In addition, the rheological  properties of linear active Brownian polymers have been studied analytically\cite{mart:18.1} and by simulations.\cite{pand:23}    

Our analytical approach reveals fundamental peculiarities of APRPs in shear flow due to their geometry and the tangential driving forces. The main result is that the stationary-state properties of the rings are independent of activity, as already obtained for APRPs in the absence of shear.\cite{phil:22} This applies to the rings' conformations as well as their rheological properties, such as the shear viscosity. This is evident by the eigenvalues of the linear but non-Hermitian eigenvalue problem, which are complex with their real parts proportional to the shear-rate-dependent relaxation times of the passive ring and the imaginary parts to activity-dependent, but shear-rate independent, frequencies. The latter determines the activity dependence of the APRPs. Considering the correlation function of the ring-diameter vector, semiflexible rings exhibit a damped periodic motion reflecting its tank-treading motion, which gradually disappears with increasing flexibility. In the presence of shear, damping increasing with increasing shear rate and the tank-treading motion becomes even less pronounced. We characterize this dynamics by a tumbling frequency, which we determine in two ways: The correlation function of the ring-diameter vector and the probability of not crossing the zero-shear plane of the $y$ component of the diameter vector. Both quantities yield a similar shear-rate dependence, particularly a crossover from tank-treading to tumbling with increasing shear rate, where the large-Weissenberg tumbling time is equal to that of a passive ring polymer.  

The article is organized as follows. Section~\ref{sec:model} introduces the semiflexible polymer model with the equation of motion and presents its solution. In Sec.~\ref{sec:conformation}, the conformational aspects of the APRP are addressed, and the dependence of the radius of gyration on the shear rate is discussed. The APRP dynamics under shear flow is discussed in Sec.~\ref{sec:dynamics} in terms of the autocorrelation function of the ring-diameter vector. A tumbling time is determined from this correlation as well as the non-crossing probability of the zero-shear plane.  Finally, Sec.~\ref{sec:summary} presents conclusions and summarizes our findings. 

\begin{figure}[t]
    \includegraphics[width = 0.9\columnwidth]{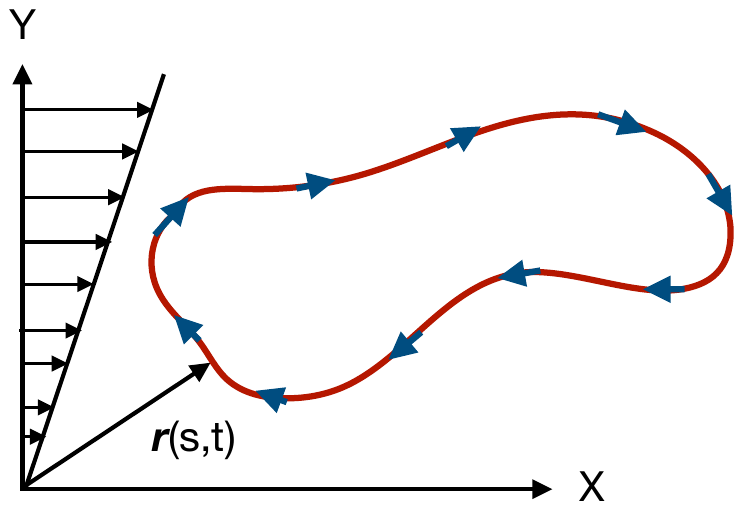}
    \caption{Illustration of a continuous active polar ring polymer (APRP) under shear flow. The arrows along its contour represent the direction of the local active force. The vector $r(s,t)$ indicates an arbitrary position on the polymer at the location $s$ and the time $t$.  The horizontal arrows illustrate the linear shear flow along the $x$ direction and the gradient direction along the $y$ direction of the Cartesian reference frame.   }  
    \label{fig:schematic}
\end{figure}
	
\section{Model} \label{sec:model}

\subsection{Equation of Motion}

A ring polymer is modeled as a continuous, differentiable space curve $\bm{r}(s,t)$ embedded in three-dimensional space, with $s \in [0,L)$ the contour variable along the ring and $t$ the time.\cite{harn:95,phil:22} Within the Gaussian semiflexible polymer model,\cite{harn:95} the  equation of motion of the point $\bm r(s,t)$ is given by \cite{phil:22}
\begin{align}  \label{eq:EOM}
	\gamma  \frac{\partial \bm{r}(s,t)}{\partial t} =  &  \  2k_BT \lambda \frac{\partial^2 \bm{r}(s,t)}{\partial s^2} - k_BT  \epsilon \frac{\partial^4\bm{r}(s,t)}{\partial s^4}  
	\nonumber
	\\ & +  f_a \frac{\partial \bm{r}(s,t)}{\partial s} +
	\gamma \mathrm{\bf K} \bm r(s,t) + \bm{\varGamma}(s,t),
\end{align}	
where $\gamma$ is the translational friction coefficient per length, $k_B$ the Boltzmann constant, and $T$ the temperature. The first and second terms on the right-hand side describe the force along the polymer contour and the bending force. The stretching coefficient  $\lambda$ (Lagrangian multiplier) accounts for the inextensibility of the ring contour and is determined by the local constraint of the mean-square tangent vector \cite{mous:19,phil:22}
\begin{equation} \label{eq:constraint}
\lla \left(\frac{\partial \bm{r}(s,t)}{\partial s} \right)^2 \rra = 1 .
\end{equation}
The bending rigidity is fixed as $\epsilon = 3/(4p)$, with $p=1/(2l_p)$ and the persistence length $l_p$. \cite{wink:03,mous:19,phil:22}   The third term accounts for the homogeneous active force of magnitude $f_a$ along the local tangent $\partial \bm{r}(s,t)/ \partial s$, and the fourth term captures the shear flow with the shear-rate tensor $\mathrm{\bf K}$. Shear is applied along the $x$ axis of the Cartesian reference frame, and the gradient direction is along the $y$ axis (see Fig.~\ref{fig:schematic}), hence, only the component $K_{xy} = \dot \gamma$ of the shear-rate tensor is different from zero, where $\dot \gamma$ is the shear rate.\cite{wink:10}      
The stochastic force $\bm{\varGamma}(s,t)$ accounts for the thermal fluctuations and is assumed to be stationary, Markovian, and Gaussian with zero mean and the second moment 
\begin{equation} \label{eq:stochastic_force}
\langle  \varGamma_{\alpha} (s,t)  \varGamma _{\beta}(s',t') \rangle = 2 \gamma k_BT  \delta_{\alpha \beta} \delta(s-s') \delta(t-t') ,
\end{equation} 
with $\alpha, \beta \in \{x,y,z \}$.

\subsection{Solution of the Equation of Motion}

Equation~\eqref{eq:EOM} is a linear but non-Hermitian Langevin equation. It can be solved by the eigenfunction expansion \cite{mous:19,phil:22}
\begin{equation} \label{eq:expansion}
\bm{r}(s,t) =\sum_{n=-\infty}^{\infty} \bm{\chi}_n(t) \phi_n(s) ,
\end{equation}
with the eigenfunctions 
\begin{equation}
\phi_n(s) = \frac{e^{i k_n s}}{\sqrt{L}} ,   
\end{equation}
the wavenumbers $k_n = 2\pi n/L$, and the normal-mode amplitudes $\bm{\chi}_n(t)$. The eigenfunctions satisfy the periodic boundary condition $\bm{r}(s,t) = \bm{r}(s+L,t)$ by the ring structure.\cite{mous:19} Insertion of the expansion \eqref{eq:expansion} into Eq.~\eqref{eq:EOM} yields the equations of motion for the normal-mode amplitudes 
\begin{equation} \label{eq:EOM_modes}
	\gamma \frac{d}{d t} \bm{\chi}_n(t) = - \xi_n \bm{\chi}_n(t) + \gamma \mathrm{\bf K} \bm \chi_n(t) + \bm{\varGamma}_n(t) .
\end{equation}	
The $\bm{\varGamma}_n(t)$ are the mode amplitudes in the eigenfunction expansion of the stochastic force $\bm{\varGamma}(s,t)$.
The eigenvalues $\xi_n$ of the eigenvalue problem are given by ($n \in \mathbb{Z}$) \cite{phil:22}
\begin{equation} \label{eq:eigenvalue}
	\xi_n = \frac{12\pi^2k_BTpL}{L^3} \left[\mu n^2 +  \frac{\pi^2}{(pL)^2}n^4 - i \frac{Pe}{6\pi pL} n \right] ,
\end{equation}	
with the abbreviation $\mu = 2 \lambda/(3p)$ and the P\'eclet number 
\begin{equation} \label{eq:pe}
	Pe = \frac{f_aL^2}{k_BT} ,
\end{equation}
which characterizes the strength of the activity.\cite{phil:22,phil:22.1,isel:15} The  non-Hermitian nature of the equation of motion \eqref{eq:EOM} yields complex eigenvalues $\xi_n$ due to presence of the first derivative. 

The complex eigenvalues $\xi_n$ are split into real and imaginary parts,
\begin{align}
\xi_n = \xi_n^R - i \xi_n^I  = \gamma/\tau_n - i \gamma \omega_n ,
\end{align}
with the relaxation times $\tau_n$ and frequencies $\omega_n$ given by
\begin{align} \label{eq:relax_time}
    \tau_n = & \ \frac{\gamma L^3}{12 \pi^2 k_B T pL} \frac{1}{\mu} \frac{1}{(n^2 + \pi^2n^4/[\mu (pL)^2])} , \\[5pt] \label{eq:frequency}
    \omega_n = & \ \frac{2 \pi f_a n }{\gamma L} = \frac{2 \pi k_BT n}{\gamma L^3} Pe.
\end{align}
Noteworthy, the relaxation times are activity independent and identical with the relaxation times of passive rings.\cite{mous:19,phil:22} Only the frequencies depend on the P\'eclet number.

\subsection{Normal-Mode Correlation Functions}

For the evaluation of expectation values, correlation functions of the normal-mode amplitudes are required. In the stationary state, the correlation functions of the Cartesian components are  ($\langle \bm \chi_n (t) \cdot \bm \chi_m (t') \rangle = \delta_{nm}  \langle \bm \chi_n (t) \cdot \bm \chi_n (t') \rangle$, $t\geq t'$, $n>0$)
\begin{align} 	\label{eq:mode_ampl} \nonumber 
	 \lla  \chi_{xn}(t)  \chi_{xn}^{*}(t') \rra  =  & \ \frac{ k_BT}{\xi_n^R} \\ &  \hspace*{-2cm } \times e^{-\xi_n (t-t')/\gamma} \left[  1 + {\dot \gamma}^2\frac{\gamma}{2 \xi_n^R} \left( t-t' + \frac{\gamma}{\xi_n^R} \right) \right] , \\  \label{eq:mode_ampl_y}
	 \lla  \chi_{yn}(t)  \chi_{yn}^{*}(t') \rra  =  & \ \frac{ k_BT}{\xi_n^R} e^{-\xi_n (t-t')/\gamma} , \\ \label{eq:mode_ampl_cross}
	 \lla  \chi_{xn}(t)  \chi_{yn}^{*}(t') \rra  =  & \ \dot \gamma \frac{k_BT}{\xi_n^R} e^{-\xi_n (t-t')/\gamma} \left( t-t' + \frac{\gamma}{2\xi_n^R} \right)  ,
\end{align}	
where $\chi_{\alpha n}^{*}(t) = \chi_{\alpha, -n}(t)$ and $\lla  \chi_{zn}(t)  \chi_{zn}^{*}(t') \rra = \lla  \chi_{yn}(t)  \chi_{yn}^{*}(t') \rra$. For the correlation functions in the passive case, see Ref.~\onlinecite{wink:10}. The tangential propulsion implies complex correlation functions with an activity-dependent frequency. This active contribution in Eq.~\eqref{eq:mode_ampl} vanishes at equal times $t=t'$, and the correlation functions in Eqs.~\eqref{eq:mode_ampl} -\eqref{eq:mode_ampl_cross} are identical to those of a  passive ring.\cite{mous:19} Hence, the conformational properties of an APRP are independent of activity and equal to those of a passive polymer \cite{phil:22} even under shear. Moreover, since the entire active force on the active ring disappears, the center-of-mass dynamics is equal to that of a passive ring. (The center of mass of the ring is given by $\bm r_{cm}(t) =  \bm \chi_0(t)/\sqrt{L}$.)
Hence, the active tangential force only affects the internal dynamic of the APRP.

\subsection{Stretching Coefficient}

The constraint for the tangent vector Eq.~\eqref{eq:constraint} leads to the equation for the calculation of the stretching coefficient $\lambda$ ($n \neq 0$)
\begin{align} \label{eq:constraint1}
L = \sum_{n = - \infty }^{\infty} k_n^2 \left[ \frac{3 k_BT}{\xi_n^R} + {\dot \gamma}^2 \frac{k_BT \gamma^2}{2 (\xi_n^R)^3} \right] .
\end{align}
Insertion of the eigenvalues \eqref{eq:eigenvalue} yields 
\begin{align}  \label{eq:constraint_wi}
1 =  \Xi(\mu) + \sum_{n=1}^\infty \frac{Wi^2}{3 pL} \frac{(\mu^0 + [\pi/(pL)]^2)^2}{n^4(\mu + [n \pi/(pL)]^2)^3},
\end{align}
where the sum in the first term on the right-hand side of Eq.~\eqref{eq:constraint1} is evaluated and yields\cite{mous:19} 
\begin{equation}
 \Xi(\mu) = \left\{ 
 \begin{array}{cc}
\displaystyle  \frac{\displaystyle 1}{\displaystyle \sqrt{\mu}} \coth(pL \sqrt{\mu})   - \frac{1}{pL \mu}  , & \mu >0 ,\\[15pt]
\displaystyle  \frac{\displaystyle 1}{\displaystyle pL |\mu|} - \frac{\displaystyle 1}{\displaystyle \sqrt{\displaystyle|\mu|}} \cot(pL \sqrt{|\mu|}) ,    &  \mu < 0 
 \end{array}
 . \right. 
\end{equation}
The Weissenberg number, $Wi$, is defined as $Wi = \dot \gamma \tau_1^0$, where $\tau_1^0$ is the longest relaxation time in Eq.~\eqref{eq:relax_time}, and $\mu^0$ is the scaled stretching coefficient both in the absence of shear. 

\begin{figure}[tb]
        \includegraphics[width=\linewidth]{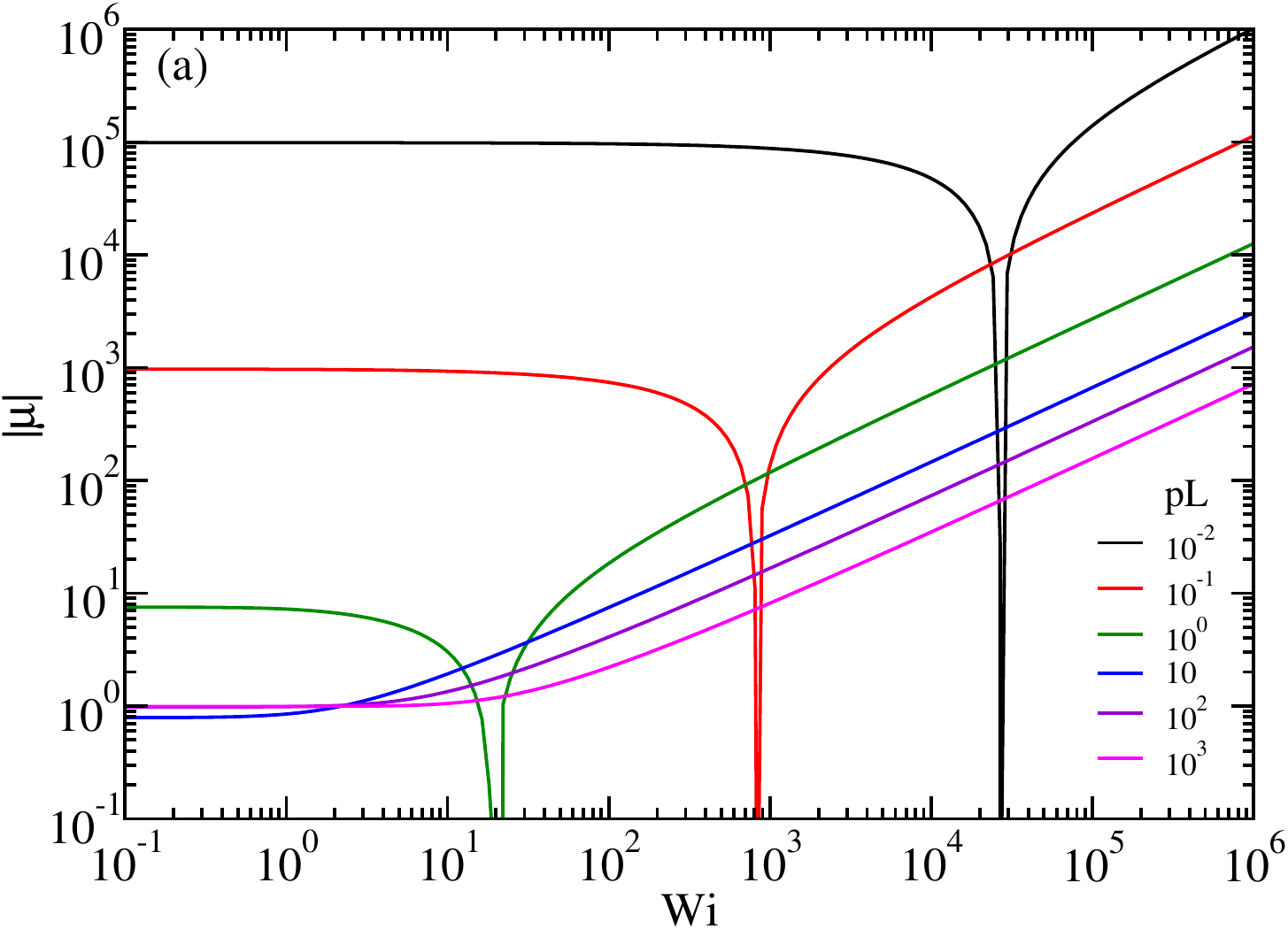} \\
        \includegraphics[width=\linewidth]{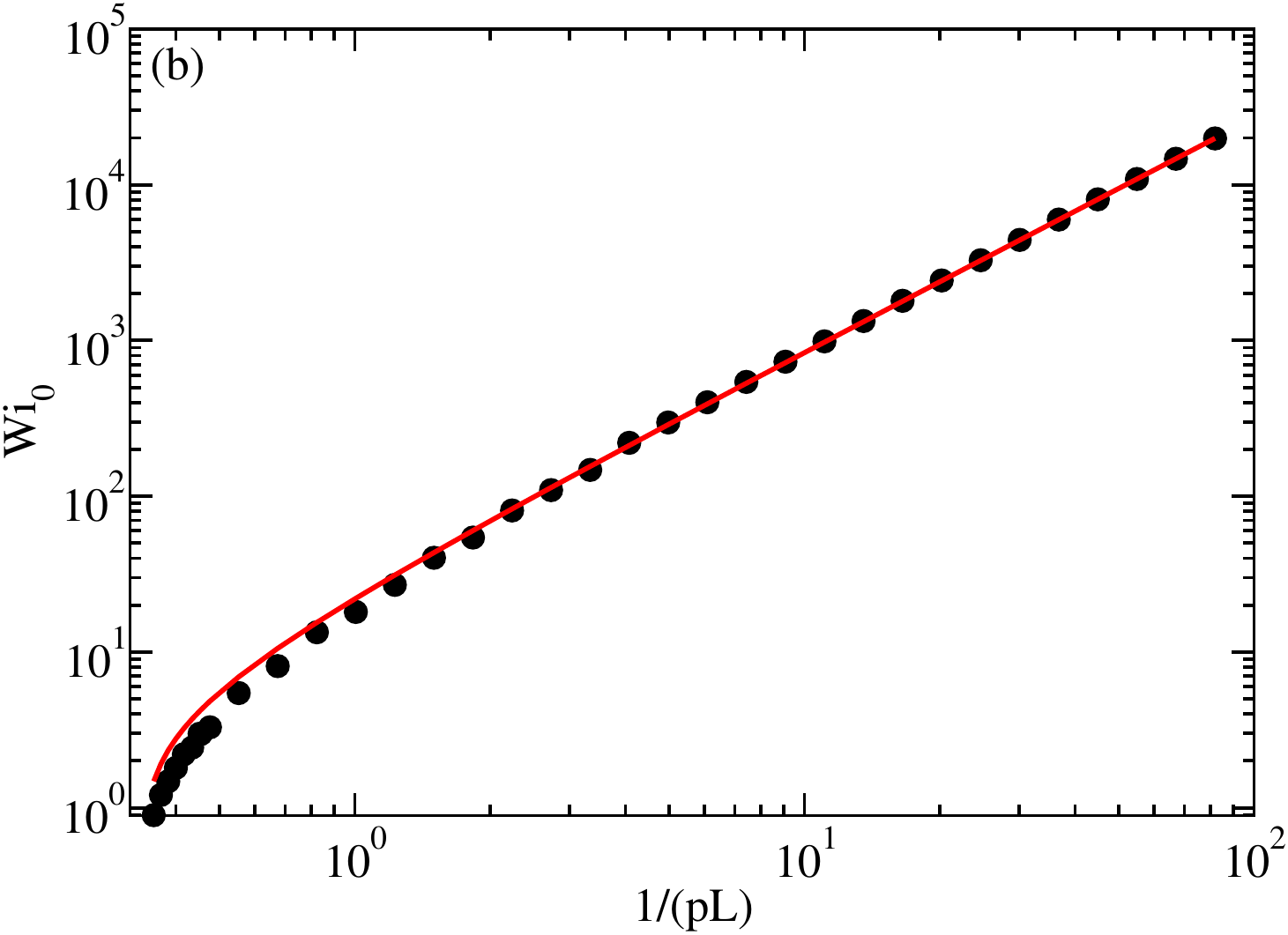}
    \caption{(a) Normalized stretching coefficient $\mu =2 \lambda /(3p)$ as a function of the Weissenberg number $Wi$ for the $pL$ values displayed in the legend. The cusp in a curve reflects the zero of the stretching coefficient and depends on the Weissenberg number. (b) Dependence of the Weissenberg number $Wi_{0}$ on the polymer stiffness $pL$ for the condition $\mu=0$.  The solid line displays the asymptotic behavior of $Wi_0$ according to Eq.~\eqref{Eq:wi_0}. }
    \label{fig:mu_wi}
\end{figure}

Figure~\ref{fig:mu_wi}(a) presents numerical solutions of Eq.~\eqref{eq:constraint_wi} for $\mu$ as a function of $Wi$ and various $pL$. For $pL > 3$, $\mu$ increases monotonically with increasing $Wi$, whereas for $pL <3$ two regimes appear, where $\mu <0$ in the range $Wi < Wi_0$ and $\mu > 0$ for $Wi> Wi_0$. The characteristic Weissenberg $Wi_0$, where $\mu = 0$, is presented in Fig.~\ref{fig:mu_wi}(b) as a function of $pL$. Asymptotically, the analytical expression
\begin{equation}
Wi_0 = \frac{\pi^3}{2 (pL)^{3/2}}   \sqrt{3- pL}
\label{Eq:wi_0}
\end{equation}
 is obtained from Eq.~\eqref{eq:constraint_wi}  in the limit $pL \ll 1$, with 
\begin{equation}
\mu^0 = -\frac{\pi^2}{(pL)^2}\left(1-\frac{2 pL}{\pi^2} \right) .    
\end{equation}
This agrees well with the numerically obtained values as shown in Fig.~\ref{fig:mu_wi}(b). The asymptotic Weissenberg number dependencies in Fig.~\ref{fig:mu_wi}(a) are given by 
\begin{equation} \label{eq:mu_approx_large_pl}
    \mu =  \left\{ 
 \begin{array}{cc}
 1 , &   Wi \ll \sqrt{3 pL}, \\[15pt]
 \displaystyle \frac{Wi^{2/3}}{\sqrt[3]{3pL}} , & Wi \gg \sqrt{3pL}  
\end{array}
\right.
\end{equation}
for $pL \gg 1$, and 
\begin{equation} \label{eq:mu_approx_small_pl}
    \mu =  \left\{ 
 \begin{array}{cc}
 \mu^0  , &  0\leq Wi \ll Wi_0 , \\[15pt]
 \displaystyle \sqrt[3]{\frac{4}{3}}\frac{Wi^{2/3}}{pL} , & Wi \gg Wi_0  
\end{array}
\right.
\end{equation}
for $pL \ll 1$. In any approximation, we only use the mode $n=1$ in Eq.~\eqref{eq:constraint_wi}.

\subsection{Relaxation Time}

The relaxation times in Eq.~\eqref{eq:relax_time} depend via $\mu$ on the shear rate. Specifically, the longest relaxation time $\tau_1$ is given by 
\begin{equation} \label{eq:tau1}
    \tau_1 =  \frac{\gamma L^3}{12 \pi^2 k_B T pL}  \frac{1}{(\mu + \pi^2/(pL)^2)} . 
\end{equation}
Its limits are  
\begin{align}
    \tau_1 = \frac{\tau_R}{4 \mu} ,  
\end{align}
for $pL \gg 1$, or $\tau_1/\tau_1^0 = 1/\mu$,
and
\begin{align}
    \tau_1 = \left\{ 
 \begin{array}{cc}
 \displaystyle  \frac{\tau_R}{8} \,  pL   , &  Wi \ll 1  \\[15pt]
\displaystyle \sqrt[3]{\frac{3}{4}} \frac{\tau_R}{4} \, pL \,  Wi^{-2/3} , & Wi  \gg 1  
\end{array}
\right. ,
\label{Eq:tau_pL}
\end{align}
for $pL \ll 1$, with the Rouse relaxation time $\tau_R = \gamma L^3/(3 \pi^2 k_BT pL)$. The scaled relaxation time $\tau_1/\tau_1^0$, where $\tau_1^0$ is the relaxation time in the absence of shear, yields in the limit $pL \ll 1$ 
\begin{align}
    \frac{\tau_1}{\tau_1^0} = \left\{ 
 \begin{array}{cc}
 \displaystyle  1  , &  Wi \ll 1  \\[15pt]
\displaystyle \sqrt[3]{6}  \,   Wi^{-2/3} , & Wi  \gg 1  
\end{array}
\right. ,
\label{Eq:tau_pL_scal}
\end{align}
i.e., a universal, stiffness, and polymer length-independent behavior is obtained. 

\begin{figure}[t]
    \includegraphics[width = \columnwidth]{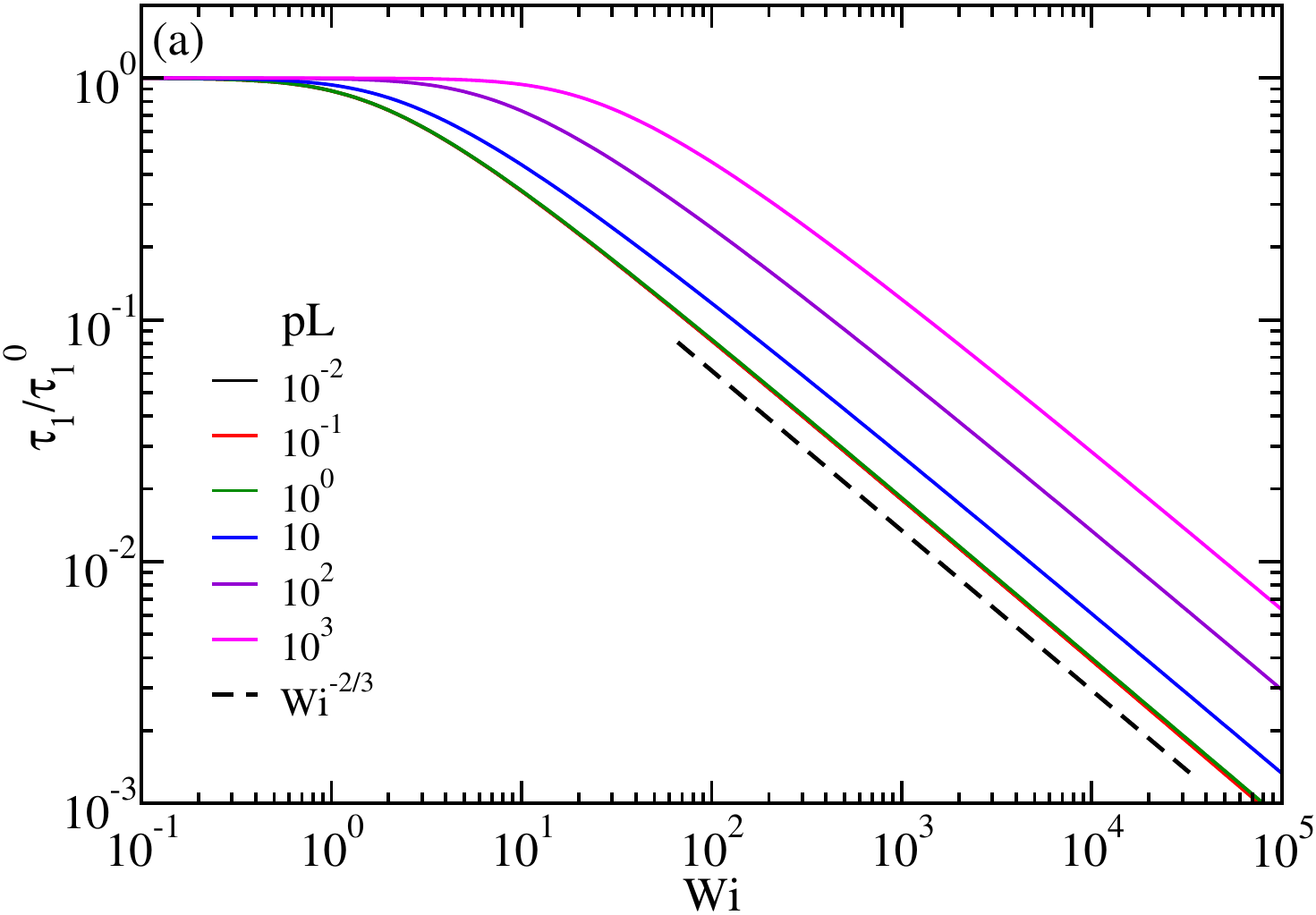}
      \includegraphics[width = \columnwidth]{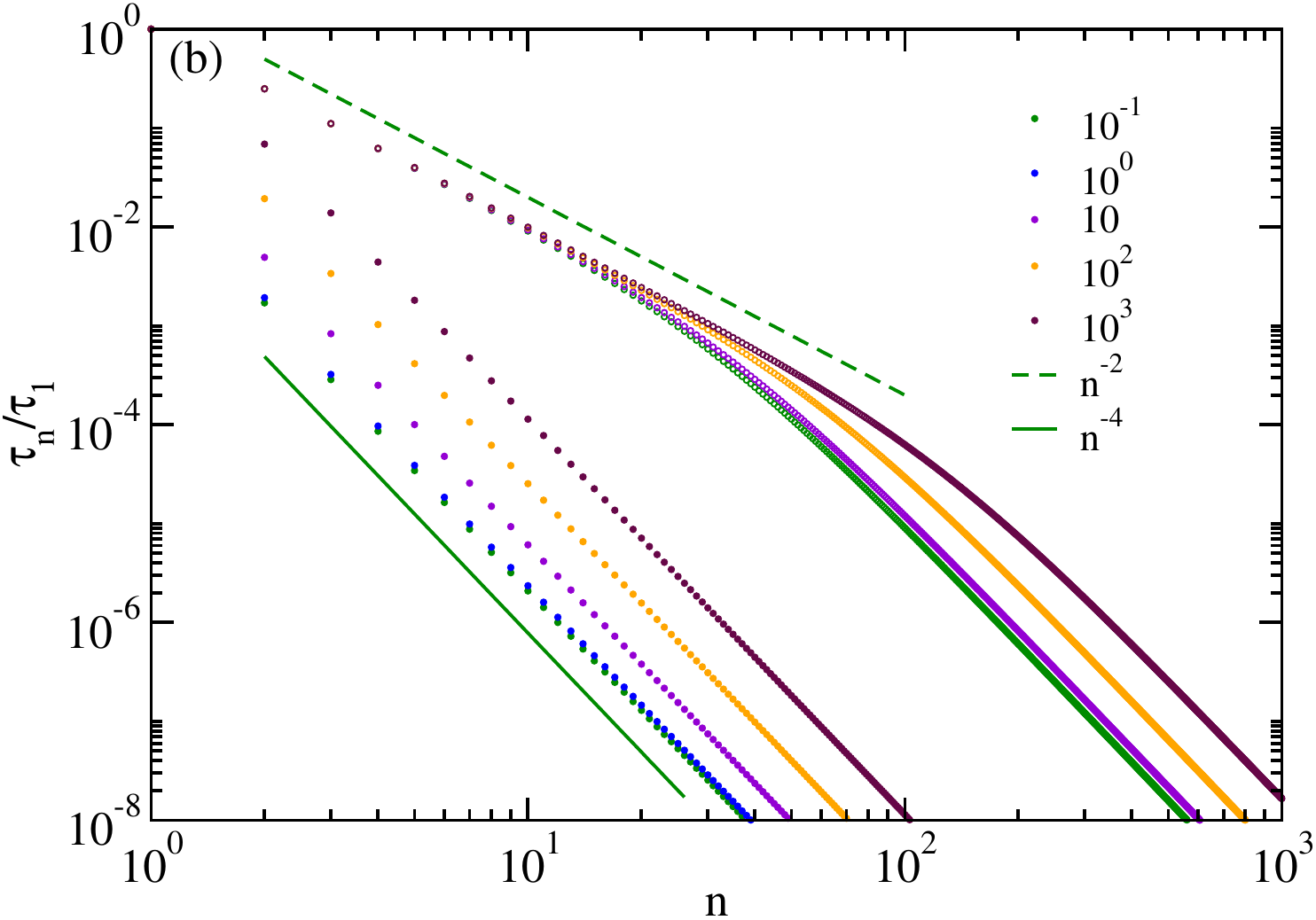}
    \caption{(a) Longest relaxation time $\tau_1$ of a ring polymer under linear shear flow as a function of the Weissenberg number $Wi$ and for various stiffnesses (legend). The scale factor $\tau_1^0$ is the longest relaxation in the absence of shear. The black dashed line shows the asymptotic power-law behavior $\tau_1\sim Wi^{-2/3}$. (b) The relaxation times $\tau_n$ as a function of the mode number $n$ for various $Wi$ (legend) at $pL=100$ (circles, top) and $pL=0.1$ (bullets, bottom). The solid line displays the $n^{-4}$ dependence of bending modes, and the dashed line the $n^{-2}$ dependence of stretching modes, respectively.\cite{harn:95}}  
    \label{fig:relaxation_time_1_wi}
\end{figure}

Figure~\ref{fig:relaxation_time_1_wi}(a)  presents normalized longest relaxation times $\tau_1/\tau_1^0$ as a function of the Weissenberg number and polymer stiffness. As predicted by Eq.~\eqref{Eq:tau_pL_scal}, the numerical values of the scaled relaxation time are independent of the ring stiffness in the limit $pL < 1$. For $pL > 3$, the $\tau_1/\tau_1^0$ curves shift toward larger Weissenberg numbers with increasing $pL$ albeit the asymptotic Weissenberg number dependence $\tau_1/\tau_1^0 \sim Wi^{-2/3}$ is  independent of $pL$. The dependence of the relaxation times $\tau_n$ on the mode number is illustrated in Fig.~\ref{fig:relaxation_time_1_wi}(b). For large stiffness, the relaxation times are dominated by bending modes ($n^{-4}$) independent of shear. In contrast, for flexible polymers, stretching modes dominate the dynamics at mode numbers $n < n_c = Wi^{1/3}(pL)^{5/6}/(3^{1/6} \pi)$, whereas a crossover occurs to a bending-mode dominated dynamics at smaller length scales ($n> n_c$). This crossover shifts to larger mode numbers with increasing shear, i.e., shear renders a ring more flexible. 

\section{Ring Conformations in Shear Flow} \label{sec:conformation}

\begin{figure}[t]
    \includegraphics[width = \columnwidth]{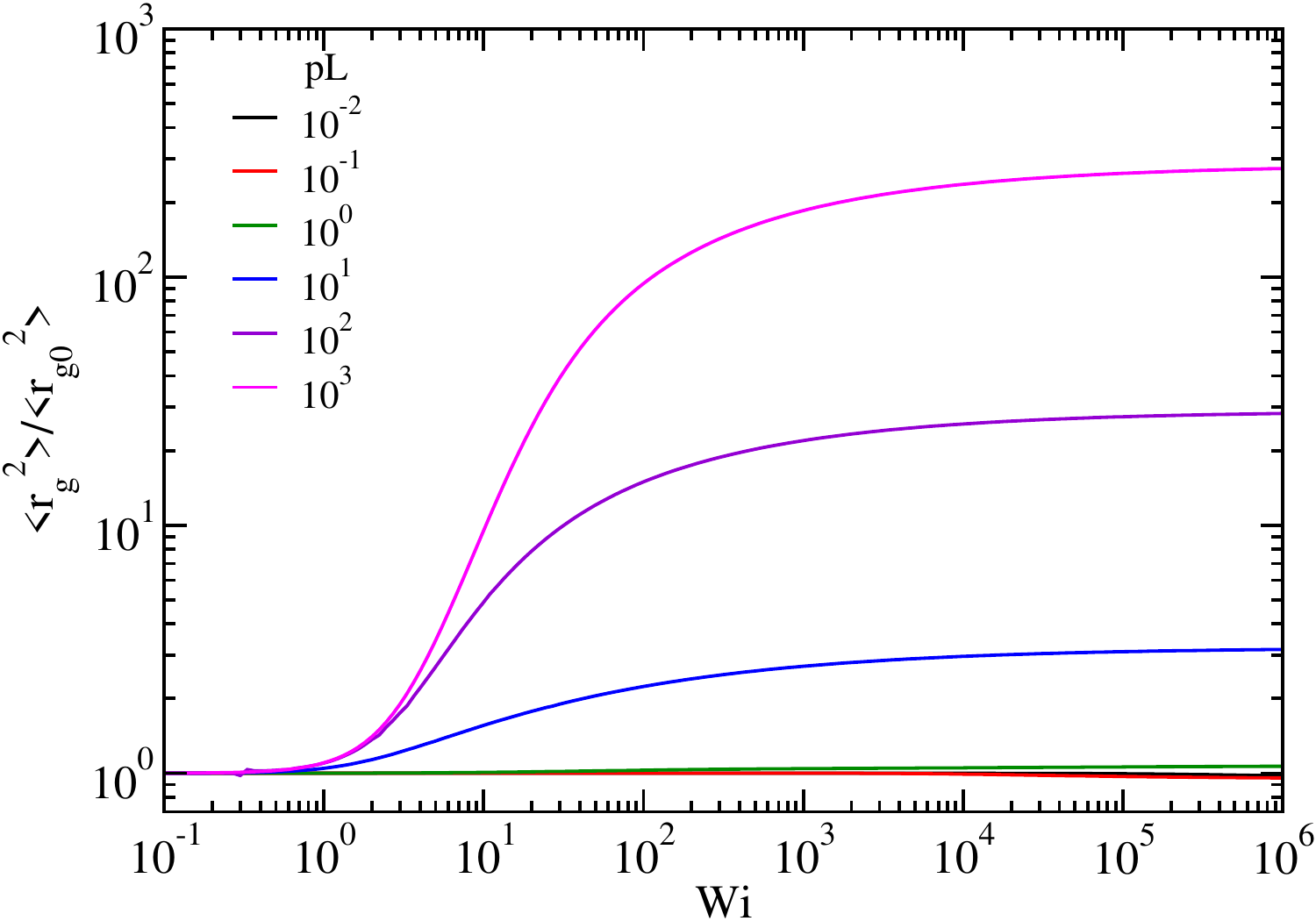}
    \caption{Normalized mean-square radius of gyration $\lla{\bm r}_g^2 \rra/\lla{\bm r}_{g0}^2\rra$ (Eq.~\eqref{eq:gyration}) of a ring polymer as a function of the Weissenberg number $Wi$ and for various stiffnesses (legend). Here, $\lla {\bm r}_{g0}^2 \rra$ is the mean-square radius of gyration in the absence of shear flow (Eq.~\eqref{eq:gyration_0}).}
    \label{fig:rg_rd}
\end{figure}

The ring conformations are characterized by its mean-square radius of gyration, 
$\langle \bm r_g^2 \rangle$, and the mean-square ring diameter, $\langle \bm r_d^2 \rangle$.\cite{mous:19}   In terms of the eigenfunction expansion \eqref{eq:expansion} the radius of gyration is given by    
\begin{align}  \label{eq:gyration} \nonumber
    {\lla {\bm r}_g^2 \rra} & \ =  \frac{1}{2L^2}\int_{0}^{L}\int_{0}^{L} \lla ({\bm r}(s)-{\bm r}(s'))^2 \rra ds ds' \  \\  
    & \ = \frac{k_BT}{L} \sum_{n=1}^{\infty} \left[ \frac{6}{\xi_{n}^R } +  \frac{\dot{\gamma}^2 \gamma^2}{(\xi_{n}^R)^3 }   \right] .
\end{align}
Substituting $\xi_n^R$ from the Eq.~\eqref{eq:eigenvalue} and inserting the Weissenberg number $Wi = \dot \gamma \tau_1^0$, we obtain
\begin{align}\label{Eq:rg_final}   \nonumber
 \lla {\bm r}_g^2 \rra  = & \ \frac{L^2}{12pL\mu}\left[ 1+\frac{3}{(pL)^2\mu}-\frac{3}{(pL)\sqrt{\mu}}\coth(pL\sqrt{\mu})\right] \\
   & \ +  \sum_{n=1}^\infty \frac{Wi^2 {L^2}}{12 \pi^2 pL} \frac{(\mu^0 + [\pi/(pL)]^2)^2}{(\mu n^2+ [ \pi n^2/(pL)]^2)^3} .
\end{align}
Similarly, we obtain the expression for the mean-square ring diameter 
\begin{align} \label{eq:msq_diamter} \nonumber 
 \lla {\bm r}_d^2 \rra = & \ \lla  ({\bm r}(L/2)-{\bm r}(0))^2 \rra \\ \nonumber 
  = & \  \frac{L^2}{4pL\mu}\left[ 1-\frac{2}{pL\sqrt{\mu}}\tanh(pL\sqrt{\mu}/2)\right] \\
 + & \ \sum_{n=1}^\infty \frac{Wi^2}{3 \pi^2 pL} \frac{{L^2}(\mu^0 + [\pi/(pL)]^2)^2}{(\mu (2n-1)^2+ [ \pi (2n-1)^2/(pL)]^2)^3}.
\end{align}
In the case $\mu <0$, analytical continuation yields $\coth(pL\sqrt{\mu}/2)/\sqrt{\mu} = -\cot(pL\sqrt{|\mu|}/2)/\sqrt{|\mu|}$ and $\tanh(pL\sqrt{\mu}/2)/\sqrt{\mu} = \tan(pL\sqrt{|\mu|}/2)/\sqrt{|\mu|}$. 
In the absence of the shear, $Wi=0$, the mean-square radius of gyration and mean-square ring diameter are given by 
\begin{equation} \label{eq:gyration_0}
  \lla \bm r_{g0}^2 \rra = \left\{  \begin{array}{cc}
 \displaystyle  \frac{L}{12 p} , & pL \gg 1  \\[10pt]
 \displaystyle  \frac{L^2}{4 \pi^2} , & pL \ll 1 
    \end{array}
    \right. ,
\end{equation}
and 
\begin{equation}
  \lla \bm r_{d0}^2 \rra = \left\{  \begin{array}{cc}
 \displaystyle  \frac{L}{4 p} , & pL \gg 1  \\[10pt]
 \displaystyle  \frac{L^2}{\pi^2} , & pL \ll 1 
    \end{array}
    \right. .
\end{equation}
Considering the first mode only in Eqs.~\eqref{eq:gyration} and \eqref{eq:msq_diamter}, the radius of gyration and the ring diameter display, up to a constant, the same shear-rate dependence, hence, for $Wi \gg 1$
\begin{align}
\lla \bm r_d^2 \rra/4 =  \lla \bm r_g^2 \rra  \approx  \frac{Wi^2 L^2}{12 \pi^2 pL} \frac{(\mu^0 + [\pi/(pL)]^2)^2}{(\mu+ [\pi/(pL)]^2)^3} . 
\end{align}
Figure \ref{fig:rg_rd} presents numerically evaluated mean-square radii of gyration as a function of the Weissenberg number and various stiffnesses. In the semiflexible limit $pL < 1$, the conformations of a ring are nearly unperturbed even for very large shear rates $Wi \gg 1$. On the contrary, in the flexible limit $pL \gg 1$, the ring displays shear-induced stretching along the flow direction upon increasing shear, which becomes more pronounced with increasing flexibility, i.e., larger $pL$. In the limit $Wi\rightarrow \infty$, the shear-rate independent ratio $\langle {\bm r}_g^2 \rangle /\langle{\bm r}_{g0}^2 \rangle \approx 3 pL/\pi^2$ is assumed, corresponding to the maximum stretching under shear. 

Transverse to the flow direction, a ring shrinks with increasing shear rate, where the shrinkage is determined by the stretching coefficient $\mu$ (first term on the right-hand side of Eq.~\eqref{Eq:rg_final}) in analogy to a linear passive polymer.\cite{wink:10}

As pointed out before, the APRP conformations are independent of activity. 

\section{Dynamics} \label{sec:dynamics}

\subsection{Tank treading} \label{ssec:tank_treading}

A polar active ring polymer exhibits an intriguing tank-treading type motion, which is most pronounced for stiff rings.\cite{phil:22} This feature is well reflected in the  autocorrelation function of the ring-diameter vector $\bm r_d(t)$ (Eq.~\eqref{eq:msq_diamter}),
\begin{align} \nonumber
& \lla \bm r_d (t) \cdot \bm r_d(0) \rra = \frac{8}{L} \sum_{n=1}^\infty \lla \bm \chi_{2n-1}(t) \cdot  \bm \chi_{2n-1}^*(0) \rra \\
& \ = \frac{8 k_B T}{\gamma L}\sum_{n=1,odd}^\infty \tau_n \left[3 + \dot \gamma^2  \tau_n (t+\tau_n)/2 \right] e^{-t/\tau_n} \cos(\omega_n t) . 
\end{align} 
The normalized correlation function in terms of the Weissenberg number reads as 
\begin{align} \nonumber \label{eq:tank_treading} \displaystyle 
 & \  C_d(t) =  \frac{\lla \bm r_d (t) \cdot \bm r_d(0) \rra }{\lla \bm r_d(0)^2 \rra}  \\
& \ = \frac{\displaystyle  \sum_{n=1,odd}^\infty \hat{\tau}_n \left[3 + Wi^2  \hat{\tau}_n^2 (1+t/\tau_n)/2  \right] e^{-t/\tau_n} \cos(\omega_n t)}{\displaystyle  \sum_{n=1,odd}^\infty \hat{\tau}_n \left[3 + Wi^2  \hat{\tau}_n^2/2 \right]  } ,
\end{align} 
with the abbreviation $\hat{\tau}_n = \tau_n/\tau_1^0$.
If we assume that the first mode dominates, the correlation function is approximately given by   
\begin{equation} \label{eq:approx_correlation}
  C_d(t)\approx  ( 1 +  t/\tau_1 )  \cos(\omega_1 t) e^{-t/\tau_1}  
\end{equation}
in the limit $\tau_1 Wi/\tau_1^0 \gg 1$. Hence, the correlation function depends only via the relaxation time $\tau_1$ on the shear rate. 

\begin{figure}[thb]
 \includegraphics[width=\columnwidth]{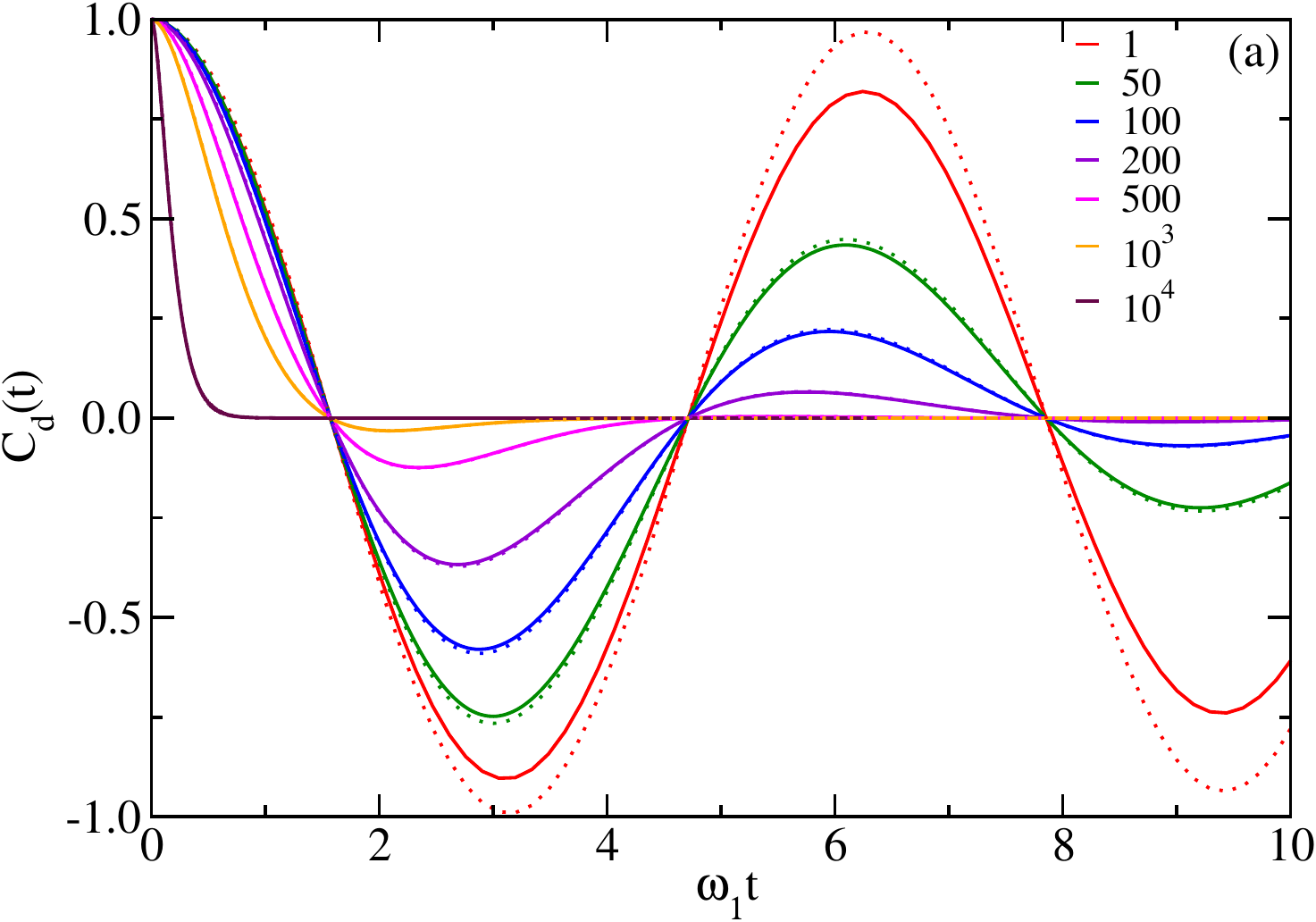}
 \includegraphics[width=\columnwidth]{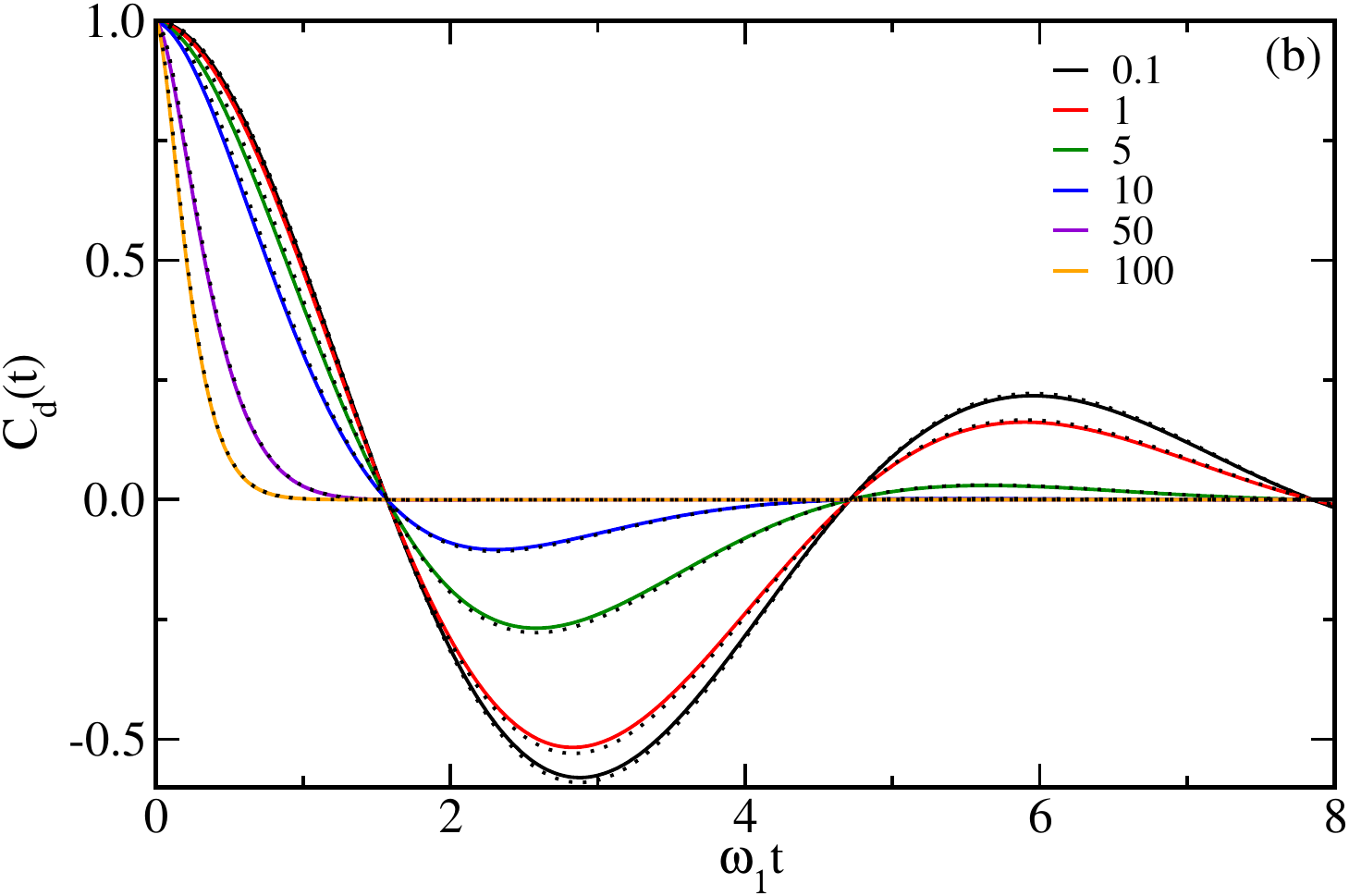}
    \caption{(a) Autocorrelation function $C_d(t)$ (Eq.~\eqref{eq:tank_treading}) of the ring-diameter vector ${\bm r}_d (t)$ as a function of the scaled time $\omega_1 t$ for various Weissenberg numbers $Wi$ (legend) and the stiffness $pL=10^{-1}$. The $\omega_1$ is the frequency of the first mode $n=1$  as given in Eq.~\eqref{eq:frequency}. 
(b) Autocorrelation function $C_d(t)$ for various  stiffnesses $pL$ (legend) and the Weissenberg number $Wi=100$.  In any case, the P{\'e}clet number is $Pe=10^3$, and the dotted lines show the approximate correlation function of Eq.~\eqref{eq:approx_correlation}. }
    \label{Fig:dia_corr}
\end{figure}

Figure~\ref{Fig:dia_corr} displays the numerically evaluated correlation function $C_d(t)$ (Eq.~\eqref{eq:tank_treading}) for various Weissenberg numbers (Fig.~\ref{Fig:dia_corr}(a)) and stiffnesses (Fig.~\ref{Fig:dia_corr}(b)). It exhibits a damped oscillatory behavior with $\omega_1 t$, signifying the tank-treading of the active ring.\cite{phil:22} As discussed in Ref.~\onlinecite{phil:22} for $Wi=0$, $C_d(t)$ exhibits a crossover to an exponential decay for $\omega_1 \tau_1 < 1$ with increasing ring flexibility $pL$. In the case $pL \ll 1$, $\omega_1 \tau_1 \approx Pe/(12 \pi)$ and $C_d(t)$ is independent of polymer length and stiffness. In the opposite limit $pL \gg1$, $\omega_1 \tau_1 \approx Pe/(6 pL)$. Thus, with increasing $pL$, the period of the oscillations is very large compared to $\tau_1$ and $C_d(t)$ decays exponentially on time scales $t/\tau_1 \lesssim 1$. 

Shear leads to a similar behavior as a function of the Weissenberg number, and oscillations are visible for $\omega_1 \tau_1 >1$, which requires $Wi$ to satisfy the condition 
\begin{equation} \label{eq:weissenberg_numb_cond}
Wi < \left\{ 
    \begin{array}{cc}
      \displaystyle   \left[Pe/(2 \pi)\right]^{3/2}/(3 pL) ,  & pL \gg 1  \\[5pt]
     \displaystyle   \left[Pe/(2 \pi)\right]^{3/2}/ 6 ,  & pL < 1 
    \end{array} 
    \right. .
\end{equation}
In the limit $pL <1$ and small $Wi$, oscillations appear independent of the ring stiffness, as for rings in the absence of shear. Here, the scaled frequency $\omega_1 \tau_1^0 = Pe/(12 \pi)$. With increasing $Wi$ and/or increasing polymer flexibility, the oscillations gradually disappear. This occurs already for rather small Weissenberg numbers at large $pL$. However, oscillations will always be present for sufficiently large P\'eclet and Weissenberg numbers satisfying Eq.~\eqref{eq:weissenberg_numb_cond}. 

The correlation $C_d(t)$ is well described by the approximation in Eq.~\eqref{eq:approx_correlation}, in particular on time scales $\omega_1 t <1$. In the limit $\tau_1 Wi/\tau_1^0 \gg 1$, the dynamics is mainly determined by the longest relaxation time. Evidently, shear implies a modified damped oscillatory behavior by the linear term $t/\tau_1$. Nevertheless, the ring dynamics is determined by two-time scales, the relaxation time $\tau_1$ and the period $2\pi/\omega_1$. In the limit $\omega_1 \tau_1 \gg 1$, the tank-trading motion is present, and for $\omega_1 \tau_1 \ll 1$, $C_d(t)$ is determined by the relaxation time $\tau_1$.   

\subsection{Tumbling}

Rather generically, polymers under shear flow exhibit a particular dynamics denoted as tumbling, where linear polymers stretch and collapse in a cyclic manner.\cite{puli:05,schr:05,wink:06.1,gera:06,delg:06,ryde:06,koba:10,huan:11,li:21.1}  More complex structures, such as star polymers\cite{sing:13} or vesicles\cite{nogu:04} perform a tank-treading type of motion, where in star polymers individual arms still undergo stretching and collapse transitions, but overall the star polymer rotates.\cite{sing:13} Ring polymers are particular and can display both modes of motion.\cite{chen:13.2,tu:20}  

In the case of passive polymers, the correlation function of the radius-of-gyration tensor components along the flow and gradient directions,\cite{huan:11} 
\begin{align}
C_{yx}(t) = \frac{\langle \Delta G_{yy}(t) \Delta G_{xx}(0) \rangle}{\sqrt{\langle \Delta G_{yy}^2\rangle \langle \Delta G_{xx}^2\rangle} } ,
\end{align}
with 
\begin{align}
G_{\alpha \alpha} = \frac{1}{L} \int_0^L \left(r_\alpha(s) - r_{\alpha}^{cm}  \right)^2 ds 
\end{align}
and $\Delta G_{\alpha \alpha} = G_{\alpha \alpha} -\langle G_{\alpha \alpha} \rangle$, 
provides a mean to calculate a tumbling time in simulations.\cite{huan:11,chen:13.2,kuma:24}  However, within our approach, this correlation function for the APRP is independent of time and, more importantly, is independent of the activity. The latter agrees with the simulation results in Ref.~\onlinecite{kuma:24}, where the extracted tumbling time corresponds to that of a passive polymer, particularly for the considered type-II activity there, which is equal to the active force in our approach. 

To decipher the tumbling dynamics of the active ring polymer in shear flow, we require a criterion that involves the frequencies $\omega_n$. As long as $\omega_1 \tau_1 \gg 1$, $\omega_1$ is the natural choice to characterize the tank-treading and tumbling motion. Yet, in the opposite case $\omega_1 \tau_1 \ll 1$, the theory predicts (practically) no periodic motion anymore, and we need another criterion. As a unifying rule, we introduce the tumbling time, $T_t$, as the time, where $C_d(t)$ has decayed to $1/e$, i.e., $C_d(T_t)=1/e$. This somewhat underestimates the tumbling time by tank-treading, but it fully captures the dependence on the P\'eclet number as long as $\omega_1 \tau_1 \gg 1$. 

\begin{figure}[t]
    \includegraphics[width=\columnwidth]{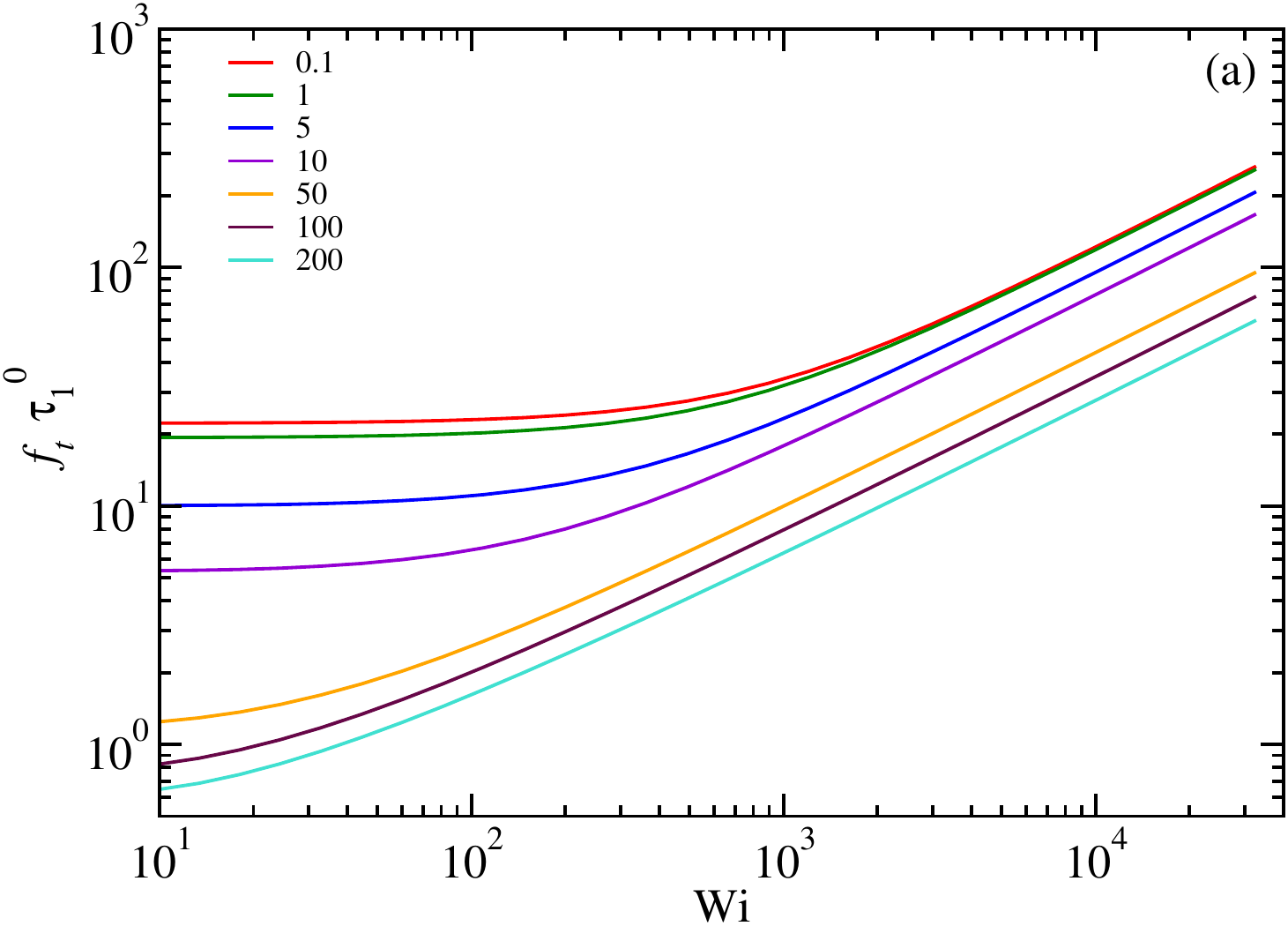}
    \includegraphics[width=\columnwidth]{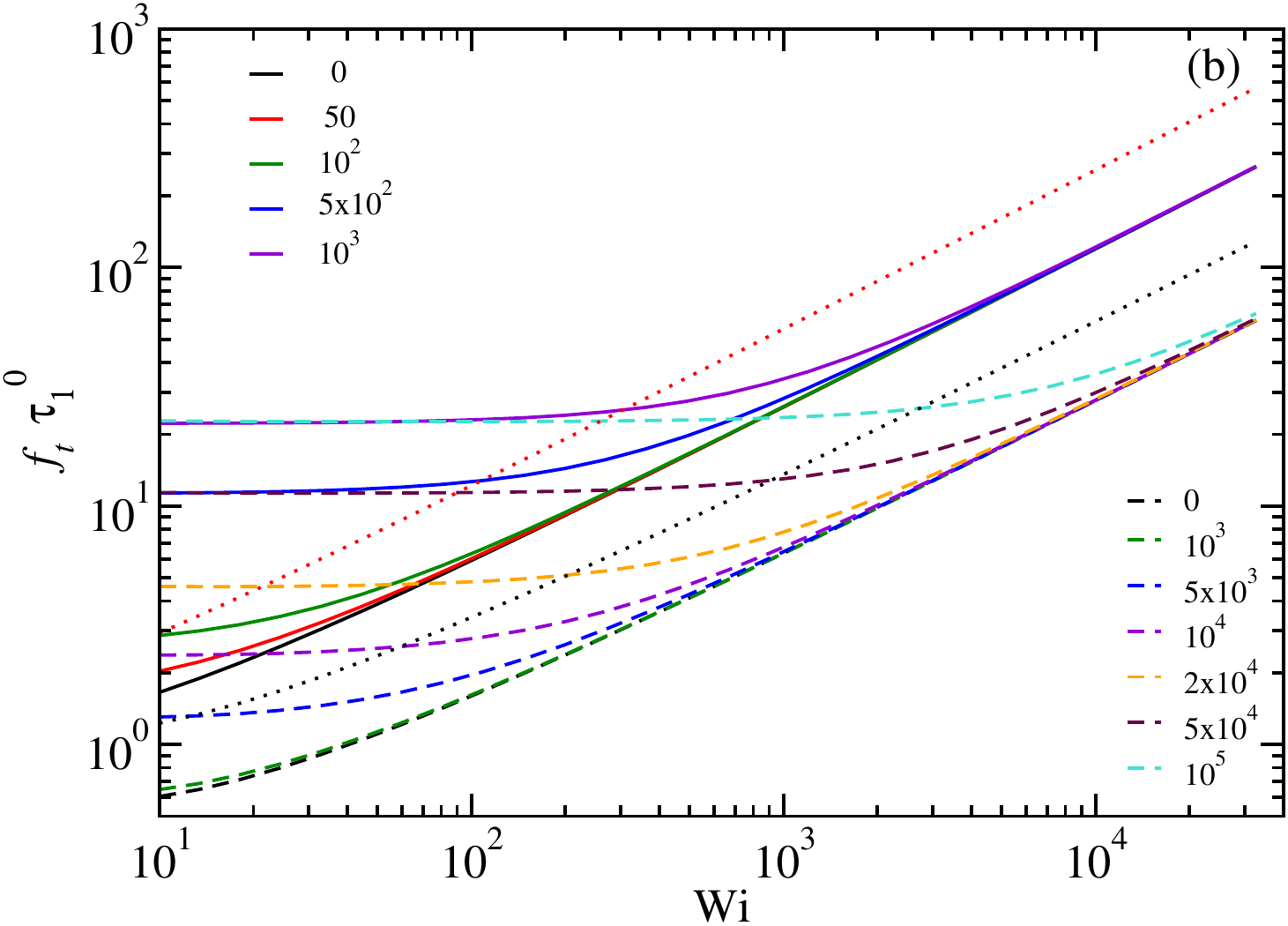}
    \caption{(a) Scaled tumbling frequency  $f_t \tau_1^0$ as a function of the Weissenberg number for various $pL$ values (legend) at P\'eclet number $Pe=10^3$. The tumbling frequency $f_t=1/T_t$ is the inverse of the time, where $C_d(T_t)=1/e$ the first time. (b) Tumbling frequency as a function of the Weissenberg number for the stiffnesses $pL =0.1$ (solid lines) and $pL=200$ (dashed lines) and for various P\'eclet numbers (legend).  The dotted lines present the ratio  $\tau_0/\tau_1$ of the largest relaxation time for $pL=0.1$ (red) and $pL=200$ (black). }
    \label{Fig:tank_frq}
\end{figure}
 
The scaled tumbling frequency $f_t = 1/T_t$ is presented in Fig.~\ref{Fig:tank_frq}(a) as a function of the Weissenberg number and various $pL$ values for $Pe=10^3$, and in Fig.~\ref{Fig:tank_frq}(b) for various P\'eclet numbers and the two values $pL=0.1$ and $pL=200$. Two regimes can be identified, specifically for stiff rings and/or large P\'eclet numbers, with a plateau at small Weissenberg numbers followed by a power-law increase at large $Wi$. A similar result has been obtained in simulations.\cite{kuma:23.1} As discussed in Sec.~\ref{ssec:tank_treading}, a ring exhibits tank-treading motion as long as the Weissenberg number obeys Eq.~\eqref{eq:weissenberg_numb_cond}. Hence, the tumbling frequency is determined by the frequency $\omega_1$, which is independent of the shear rate. Moreover, $\omega_1 \tau_1^0$ is independent of polymer stiffness and length for $pL \ll 1$, which is reflected in the similarity of the plateau values for $pL <1$ in Fig.~\ref{Fig:tank_frq}(a), but the product depends linearly on $Pe$ (Eq.~\eqref{eq:frequency}). Hence, the plateau value decreases linearly with increasing $Pe$ as can be discerned from Fig.~\ref{Fig:tank_frq}(b). With increasing Weissenberg number, the oscillation in $C_d(t)$ gradually disappears, and the tumbling time is determined by the modified exponential function including the linear term (Eq.~\eqref{eq:approx_correlation}). Neglecting the cosine term in Eq.~\eqref{eq:approx_correlation}, the tumbling time following from the condition $C_d(T_t) = 1/e$ is given by $T_t/\tau_1 \approx 2.1$, or $f_t \tau_1^0=\tau_1^0/(2.1 \tau_1)$. Thus, in the limit $pL \gg 1$, $f_t \tau_1^0=\mu/2.1$. The tumbling time is independent of $Pe$, as shown in Fig.~\ref{Fig:tank_frq}(b), and increases as $Wi^{2/3}$ with increasing Weissenberg number. Moreover, $\mu$ -- hence $f_t$ -- decreases with increasing $pL$ (Eq.~\eqref{eq:mu_approx_large_pl}) as reflected in Fig.~\ref{Fig:tank_frq} and Fig.~\ref{fig:mu_wi}(a). The dotted lines in Fig.~\ref{Fig:tank_frq}(b) indicate $\tau_1^0/\tau_1$, which corresponds to a tumbling time following an exponential decay only. In the limit $Wi \gg 1$, the longest relaxation time captures the dependence on the Weissenberg number, as is the case of passive linear polymers under shear flow,\cite{wink:06.1,huan:11} however, it deviates quantitatively by approximately the factor $2.1$ due to the linear time-dependent term in the correlation function $C_d(t)$.  

The scaled tumbling frequencies for $pL=0.1$, $Pe \geq 5\times 10^2$ and $pL=200$, $Pe \geq 5\times 10^4$  coincidentally coincide in the limit $Wi < 10$ (Fig.~\ref{Fig:tank_frq}(b)). This follows from the definition of $\omega_1$ and the approximations of $\tau_1^0$ in the limits $pL \ll 1$ and $pL \gg 1$, which yields $\omega_1 \tau_1^0 = Pe/(12 \pi)$ for $pL=0.1$ and $\omega_1 \tau_1^0 = Pe/(6 \pi pL)$ for $pL \gg 1$. Insertion of the $Pe$ and $pL$ values yields identical products.

We would like to point out that the correlation function $C_d(t)$ applied here to calculate a tumbling time is equivalent to the correlation function $C_x$ introduced in Ref.~\onlinecite{kuma:24}. Calculating the correlation function $C_x$ within the current model yields exactly the same shear-rate dependence as Eq.~\eqref{eq:tank_treading}, only the shear-independent term differs by a factor three due to the considered three-dimensional correlation rather than a one-dimensional one.

\subsection{Probability for crossing zero-shear plane}

The tumbling dynamics of the ring can also be characterized by the probability $\psi(r_{yd}^0,t)$ of the $y$ component of the radius vector, $r_{yd}$, of not crossing the zero-shear plane ($r_{yd} =0$) up to the time $t$, with the initial position $r_{yd}^0$ at the time $t=0$.\cite{wink:06.1,maju:96} Assuming that the first mode dominates the ring dynamics, $\bm r_d$ is given by
\begin{equation}
   \bm r_d(t) = \frac{2}{\sqrt{L}} \left( \bm \chi_1 + \bm \chi_1^*\right) = \frac{4}{\sqrt{L}} \bm \chi_1^R ,   
\end{equation}
with $\bm \chi_1^R$ the real part of $\bm \chi_1$, or $r_{yd} = 4 \chi_{y 1}^R/ \sqrt{L}$. The mode amplitudes $\chi_{y1}$ and $\chi_{y1}^*$ obey the linear Langevin equation~\eqref{eq:EOM_modes}, i.e., correspond to an Ornstein-Uhlenbeck process.\cite{risk:89,wink:06.1} Moreover, the equations of motion do not explicitly depend on the shear rate. The  solution of the coupled Langevin equations is presented in Appendix~\ref{app:tumbling}. The probability of not crossing the zero shear plane up to the time $t$  is given by (Eq.~\eqref{app:probability_cross})
\begin{align}  \label{eq:probability_cross}
    \psi(r_{yd}^0,t) = \mathrm{erf} \left( \frac{\sqrt{L} r_{yd}^0 \, e^{-t/\tau_1} \cos(\omega_1 t) }{4 \sqrt{k_BT \tau_1 (1-e^{-2 t/\tau_1})/\gamma}} \right) ,
\end{align}
where $r_{yd}^0$ is the (positive) value of the diameter-vector component at the time $t=0$. 
Taylor expansion of the error function for small arguments yields
\begin{align}
\psi(r_{yd}^0,t) \approx \frac{\sqrt{L} r_{yd}^0 \, e^{-t/\tau_1} \cos(\omega_1 t) }{2 \sqrt{k_BT \tau_1 \pi  (1-e^{-2 t/\tau_1})/\gamma}} .
\end{align}
Hence, the distribution function exhibits damped oscillations for $t/\tau_1 > 1/2$ as long as the first mode dominates and $Wi \gg 1$. In particular, it does not include a linear term as the correlation function of Eq.~\eqref{eq:approx_correlation}. 

\begin{figure}[t]
    \includegraphics[width =\columnwidth]{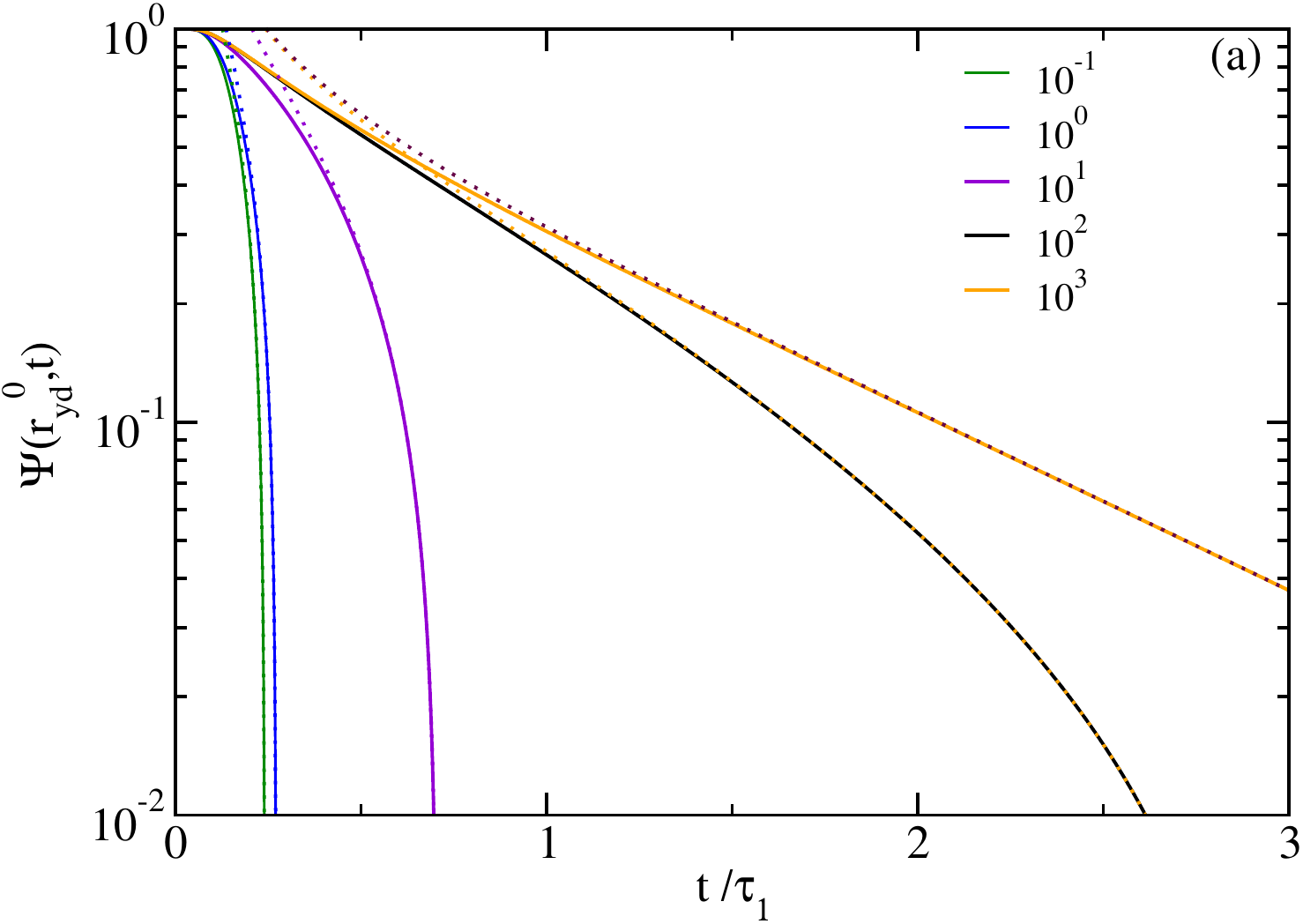}
    \includegraphics[width = \columnwidth]{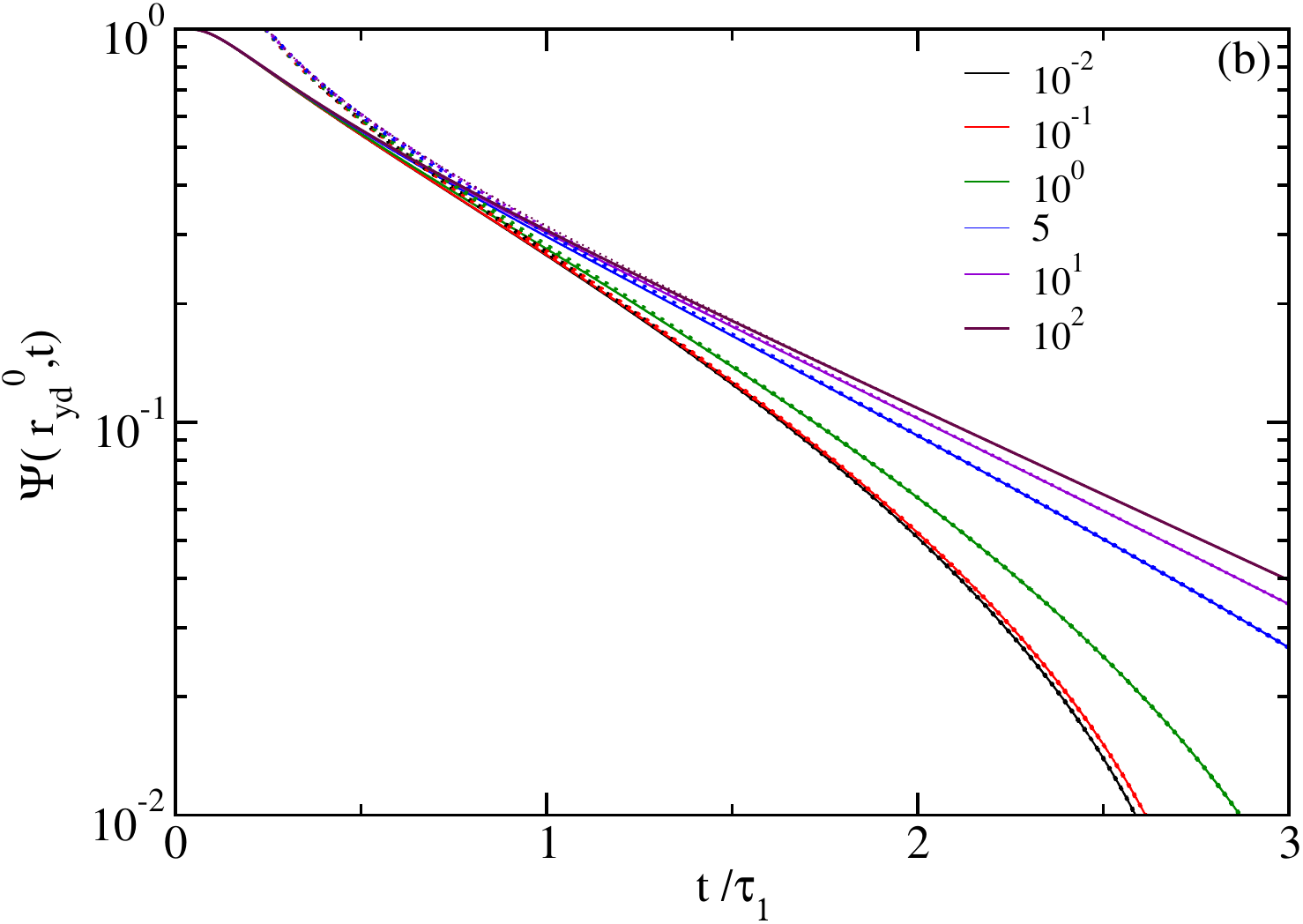}
    \caption{Probability distribution function $\psi(r_{yd}^0,t)$ of $r_{yd}$ of not crossing the zero shear plane up to the time $t$ as a function of the scaled time $t/\tau_1$ for (a) $pL=0.1$ and various Weissenberg numbers (legend), and (b) various $pL$ values (legend) for the  Weissenberg number $Wi=10^2$. The P\'eclet number is $Pe =10^3$. The dashed lines show the approximation for $\psi(r_{yd}^0,t)$ according to  Eq.~\eqref{eq:probability_approx}. }
    \label{fig:phi_dis}
\end{figure}
 
Choosing for $r_{yd}^0$ the root-mean-square value of the diameter vector in the $y$ direction of a passive ring under shear flow and considering only the first mode, yields
\begin{equation}
r_{yd}^0 = \sqrt{\langle r_{yd}^2 \rangle} = \sqrt{8 k_BT \tau_1/(\gamma L)}
\end{equation}
and
\begin{align}  \label{eq:probability_cross_1}
    \psi(r_{yd}^0,t) = \mathrm{erf} \left( \frac{ e^{-t/\tau_1} \cos(\omega_1 t) }{\sqrt{2}  (1-e^{-2 t/\tau_1})} \right) .
\end{align}
Taylor expansion of the error function for small arguments results in
\begin{align}  \label{eq:probability_approx}
    \psi(r_{yd}^0,t) \approx \sqrt{\frac{2}{\pi}}\frac{ e^{-t/\tau_1} \cos(\omega_1 t) }{\sqrt{  (1-e^{-2 t/\tau_1})}}.
\end{align}

Figure~\ref{fig:phi_dis} provides examples of the distribution function for various Weissenberg numbers and stiffnesses. As for the correlation function $C_d(t)$, two regimes are present, an exponentially decaying regime for $\omega_1 \tau_1 < 1$ and a regime dominated by the cosine term for $\omega_1 \tau_1 > 1$. Since the approximation in Eq.~\eqref{eq:probability_approx} describes the distribution function very well for $t/\tau_1>1/2$, this approximation can characterize the tumbling dynamics. The main difference to Eq.~\eqref{eq:approx_correlation} is the absence of the linear time-dependent term. Focusing on the term $e^{-t/\tau_1} \cos (\omega_1 t)$ only and applying the same rule as for the correlation function $C_d(t)$, namely a decay of the quantity to $1/e$, we obtain the same features for the tumbling frequency as in Fig.~\ref{Fig:tank_frq}, namely a plateau for $\omega_1 \tau_1 >1$ and a power-law increase for $\omega_1 \tau_1 < 1$. However, in the latter case, the tumbling time is given by $T_t=\tau_1$ as for passive linear polymers.\cite{wink:06.1,huan:11} Hence, the tumbling of a ring polymer can be characterized by the real part of the correlation function in Eq.~\eqref{eq:mode_ampl_y} for the first mode:\cite{wink:06.1,maju:96} 
\begin{equation}
   \mathrm{Re} \left( \frac{\lla \chi_{y1}(t) \chi^*_{y1}(0) \rra}{|\chi_{y1}(0)|^2} \right)  = e^{-t/\tau} \cos(\omega_1 t).
\end{equation}

\section{Summary and Conclusions} \label{sec:summary}

In this article, we have presented analytical results for the conformational and dynamical properties of  active polar ring polymers (APRPs) subjected to linear shear flow. The active force is imposed along the local tangent of the continuous Gaussian semiflexible phantom polymer.\cite{phil:22} The solution of the linear non-Hermitian  equation of motion is obtained by an eigenfunction expansion. The respective eigenvalues are complex, with shear rate-dependent, activity-independent relaxation times (real part) and activity-dependent, shear rate-independent frequencies (imaginary part). The relaxation times are those of passive ring polymers under shear flow. As a consequence, the ring polymer conformations, which are independent of the frequencies, are independent of activity and are identical to those of passive ring polymers in shear.\cite{phil:22} Hence, the APRP's deformations and rheological properties depend solely on shear.  

Yet, the ring dynamics is governed by the activity as long as the product of the smallest frequency and longest relaxation time $\omega_1 \tau_1 >1$. Then, a ring exhibits a tank-treading motion, as is reflected in oscillations of the correlation function $C_d(t)$ of the ring-diameter vector.\cite{phil:22} As long as the Weissenberg number obeys the condition of Eq.~\eqref{eq:weissenberg_numb_cond}, $C_d(t)$ is independent of stiffness for $pL<1$. With increasing Weissenberg number, the product $\omega_1 \tau_1 <1$ independent of stiffness and relaxation dominates over oscillations, which implies a modified exponential decay of $C_d(t)$ (Eq.~\eqref{eq:approx_correlation}). 

The presence of two characteristic time scales -- the longest relaxation time $\tau_1$ and the frequency $\omega_1$ -- is reflected in the tumbling dynamics of a ring polymer, with a tank-treading motion for $\omega_1 \tau_1 >1$ (Eq.~\eqref{eq:weissenberg_numb_cond}) and strong conformational changes in the opposite case, where relaxation dominates. Hence, an increase of the Weissenberg number implies a softening of an originally stiff polymer. The tumbling frequency exhibits a plateau at small shear rates and a power-law increase at large Weissenberg numbers. The plateau value is determined by the frequency $\omega_1$ and is proportional to the P\'eclet number, whereas the power-law increase at large $Wi$ is governed by the longest relaxation time $\tau_1 \sim Wi^{-2/3}$.\cite{wink:06.1,huan:11} A similar crossover in the tumbling frequency has been obtained in Ref.~\onlinecite{kuma:24}.          

The tumbling dynamics of ring polymers can be characterized by various quantities. We have chosen two quantities, the ring-diameter correlation function $C_d(t)$ and the probability $\psi$ of not crossing the zero-shear plane up to the time $t$ of the $y$ component of the diameter vector. Both quantities yield qualitatively the same shear-rate dependence. However, $C_d(t)$ exhibits a modified exponential decay, including a linear time-dependent term. Qualitatively, this does not affect the power-law dependence of the tumbling frequency on the shear rate in our case, but may do so if another criterion is applied. The linear term is absent in the probability $\psi$. Hence, we consider this quantity better suited to characterize the tumbling dynamics, particularly in the limit of large shear rates.   

The way how an active force on a polymer is implemented can substantially affect its behavior. This applies to polar polymers, where differences in the choice of the tangential force in a discrete model lead to fundamental differences in a polymer's conformational properties,\cite{bian:18,anan:18,loca:21,phil:22.1,faze:23} as well as to other kinds of active forces such as in active Brownian polymers (ABPOs).\cite{eise:16,anan:20,wink:20} In addition, excluded-volume effects contribute to conformational changes. Naturally, such effects also influence the polymer's properties in shear flow.\cite{mart:18.1,pand:23,kuma:24} For active polar ring polymers under shear, different bond potentials have been considered and differences revealed.\cite{kuma:24} Hence, a model for a comparison with experimental results must be carefully and appropriately selected. We consider the presented approach rather generic, which should capture essential features of APRPs in shear flow over a wide range of P\'eclet numbers. Thus, we expect it to be helpful in the interpretation of observations on ring polymers under shear.

\section*{AUTHOR DECLARATIONS}

\section*{Conflict of Interest}

The authors have no conflicts of interest to declare.

\section*{Data Availability Statement}

The data that support the findings of this study are available from the corresponding author upon reasonable request.

\section*{Author Contributions}

{\bf Roland G. Winkler:} Conceptualization (equal); Project administration (lead); Formal Analysis (lead); Writing – original draft (lead); Writing – review \& editing (equal).

{\bf Sunil P. Singh:} Conceptualization (equal); Formal Analysis (support); Software (lead); Visualization (lead); Writing – review \& editing (equal).

\section*{Acknowledgements}
SPS acknowledges funding support from the DST-SERB Grant No. CRG/2020/000661, computational time at IISER Bhopal, and useful discussion with  S. Thakur. 

\appendix

\section{Probability Distribution} \label{app:tumbling}

The Langevin equations for the normal-mode components $\chi_{1y}$ and $\chi_{1y}^*$ read
\begin{align}
   \gamma \dot \chi_{1y}(t) & \ = - \xi_1 \chi_{1y} + \varGamma_{1y} , \\
     \gamma \dot \chi_{1y}^*(t) & \ = - \xi_1^* \chi_{1y}^* + \varGamma_{1y}^* .
\end{align}
Suppressing the indices ($1y$) and separating $\chi_{1y}$,  $\chi_{1y}^*$ into their real and imaginary parts, $\chi_{1y} = \chi^R + i \chi^I$, the equations of motion of the latter are
\begin{align} \label{app:equation_motion}
   \dot{\tilde{\bm \chi}}(t) = \mathrm{\bf A} \tilde {\bm \chi}(t) + \frac{1}{\gamma} \tilde{\bm \varGamma}(t), 
\end{align}
with $\tilde{\bm \chi} = (\chi^R,\chi^I)^T$, $\tilde{\bm \varGamma} = (\varGamma^R, \varGamma^I)^T$, and 
\begin{equation} \displaystyle 
\mathrm{\bf A} = \left( \begin{array}{cc}  \displaystyle 
 - \frac{1}{\tau} & - \omega \\ 
\omega & -  \displaystyle  \frac{1}{\tau} \end{array} \right) .
\end{equation}
Note that the index "1" is suppressed in the relaxation time and the frequency. The calculation of the correlation functions of the stochastic forces yields 
\begin{align}
    \lla \varGamma^R(t) \varGamma^R(t') \rra  & \ = \lla \varGamma^I(t) \varGamma^I(t') \rra = \gamma k_B T \delta(t-t') , \\
     \lla \varGamma^R(t) \varGamma^I(t') \rra & \ = 0.
\end{align}
The solution of the Eq.~\eqref{app:equation_motion} is given by
\begin{equation} 
     \tilde{\bm \chi}(t) = \lla \tilde{\bm \chi}(t) \rra  + \frac{1}{\gamma} \int_0^t dt' e^{-(t-t')/\tau} \mathrm{\bf G}(t-t') \tilde{\bm \varGamma}(t') \, dt' , 
\end{equation}
with the matrix
\begin{align}
\mathrm{\bf G}(t) = 
    \left( \begin{array}{cc} 
    \cos(\omega t) & - \sin(\omega t) \\
    \sin(\omega t) & \cos(\omega t)
    \end{array} \right) , 
\end{align}
the average 
\begin{equation}
   \lla \tilde{\bm \chi}(t) \rra = e^{-t/\tau}   \mathrm{\bf G}(t) \tilde{\bm \chi}^0 , 
\end{equation}
and the initial condition $\tilde{\bm \chi}^0 = (\chi^R_0, \chi^I_0)^T = (\chi^R(0), \chi^I(0))^T$. In the following, we consider the initial condition $\chi^I(0)=0$, hence, 
\begin{equation}
    \lla \chi^R(t)\rra = \chi^R_0  e^{-t/\tau}  \cos(\omega t) .
\end{equation}
Calculation  of the second moment yields
\begin{align}
    & \ \sigma_{RR} = \lla (\chi^R(t) - \lla \chi^R\rra)^2 \rra  =   \frac{k_BT \tau}{2 \gamma} \left(1-e^{-2t/\tau} \right) , \\
   & \ \sigma_{II} =  \lla (\chi^I(t) - \lla \chi^I \rra)^2 \rra = \lla (\chi^R(t) - \lla \chi^R\rra)^2 \rra   , \\[5pt]
    & \ \sigma_{RI} = \sigma_{IR} = \lla (\chi^R(t) - \lla \chi^R\rra) (\chi^I(t) - \lla \chi^I \rra) \rra  = 0 .  
\end{align}
The distribution function $\psi(\chi^R,t|\chi^R_0,0)$, which follows as a solution of the Fokker-Planck equation corresponding to the Langevin equation~\eqref{app:equation_motion} and which satisfies the absorbing boundary condition $\psi(0,t|\chi^R_0,0) =0$, is then given by \cite{risk:89}
\begin{align} \nonumber 
 &    \psi(\chi^R,t|\chi^R_0,0) = \frac{1}{\sqrt{2\pi \sigma_{xx}}} \\
 & \hspace*{1cm} \times \left(  e^{- (\chi^R-\langle \chi^R \rangle)^2/(2 \sigma_{RR}) } - e^{- (\chi^R+\langle \chi^R \rangle)^2/(2 \sigma_{RR}) }\right).
\end{align}
The probability of not crossing the zero-shear plane up to the time $t$ is then obtained as 
\begin{align} \nonumber \label{app:probability_cross}
    \psi(\chi^R_0,t) & \ = \int_0^{\infty} \psi(\chi^R,t|\chi^R_0,0) \, d \chi^R  \\ 
    & \ = \mathrm{erf} (\langle \chi^R\rangle /\sqrt{2 \sigma_{RR}}) .
\end{align}

%\bibliography{bibliography}
%\bibliography{/Users/winkler/ownCloud/publications_library/bibliography/bibliography}
%merlin.mbs aipnum4-1.bst 2010-07-25 4.21a (PWD, AO, DPC) hacked
%Control: key (0)
%Control: author (8) initials jnrlst
%Control: editor formatted (1) identically to author
%Control: production of article title (-1) disabled
%Control: page (0) single
%Control: year (1) truncated
%Control: production of eprint (0) enabled
%

\end{document}